\documentclass[apjl]{emulateapj}
\usepackage{apjfonts}

\shorttitle{STRUCTURE AND HISTORY OF DM HALOS PROBED WITH GL}
\shortauthors{LAPI \& CAVALIERE}
\journalinfo{Accepted by ApJ Letters.}

\begin{document}

\title{Structure and History of Dark Matter Halos Probed with Gravitational Lensing}
\author{A. Lapi\altaffilmark{1,2} and A. Cavaliere\altaffilmark{1,3}}
\altaffiltext{1}{Astrofisica, Dip. Fisica, Univ. `Tor Vergata',
via Ricerca Scientifica 1, 00133 Roma, Italy.}
\altaffiltext{2}{Astrophysics Sector, SISSA/ISAS, Via Beirut
2-4, 34014 Trieste, Italy.} \altaffiltext{3}{Accademia dei
Lincei, Via Lungara 10, 00165 Roma, Italy.}

\begin{abstract}
We test with gravitational lensing data the dark matter (DM)
halos embedding the luminous baryonic component of galaxy
clusters; our benchmark is provided by their two-stage
cosmogonical development that we compute with its variance, and
by the related `$\alpha$-profiles' we derive. The latter solve
the Jeans equation for the self-gravitating, anisotropic DM
equilibria, and yield the radial runs of the density $\rho(r)$
and the velocity dispersion $\sigma_r^2(r)$ in terms of the DM
`entropy' $K \equiv \sigma_r^2/\rho^{2/3}\propto r^{\alpha}$
highlighted by recent $N$-body simulations; the former
constrains the slope to the narrow range $\alpha\approx 1.25 -
1.3$. These physically based $\alpha$-profiles meet the overall
requirements from gravitational lensing observations, being
intrinsically flatter at the center and steeper in the
outskirts relative to the empirical NFW formula. Specifically,
we project them along the l.o.s. and compare with a recent
extensive dataset from strong and weak lensing observations in
and around the cluster A1689. We find an optimal fit at both
small and large scales in terms of a halo constituted by an
early body with $\alpha\approx 1.25$ and by recent extensive
outskirts, that make up an overall mass $10^{15}\, M_{\odot}$
with a concentration parameter $c\approx 10$ consistent with
the variance we compute in the $\Lambda$CDM cosmogony. The
resulting structure corresponds to a potential well shallow in
the outskirts as that inferred from the X rays radiated from
the hot electrons and baryons constituting the intracluster
plasma.
\end{abstract}


\section{Introduction}

The collisionless, cold dark matter (DM) particles that
constitute the gravitationally dominant component of galaxy
clusters are distributed in a `halo' embedding the
electromagnetically active baryons. The halo development under
self-gravity from small density perturbations has been focused
by several recent $N$-body simulations, with three main
outcomes.

First, the growth is recognized (Zhao et al. 2003; Hoffman et
al. 2007; Diemand et al. 2007) to comprise two cosmogonic
stages: an early fast collapse including a few violent major
mergers building up the halo `body'; a later, quasi-equilibrium
stage where the outskirts develop from the inside-out by minor
mergers and smooth accretion (see Salvador-Sol\'e et al. 2007).
The \emph{transition} occurs at the redshift $z_t$ when a DM
gravitational well attains its maximal depth, or the radial
peak of the circular velocity $v^2_c\equiv G M/R$ its maximal
height (see Li et al. 2007). This sharp definition of halo
`formation' also marks the time for the early gravitational
turmoil to subside.

Second, the ensuing quasi-equilibrium structure is effectively
expressed in terms of the functional $K\equiv
\sigma_r^2/\rho^{2/3}$ that combines the density $\rho$ and the
radial velocity dispersion $\sigma_r^2$ in the form of a DM
`entropy' (or rather `adiabat'; see Bertschinger 1985, Taylor
\& Navarro 2001, Hoffman et al. 2007, Vass et al. 2008). This
mimics the behavior of a thermodynamic entropy in that it
increases in the halo bodies during the fast collapses and
stays put during the subsequent quiet accretion.

Third, the simple run $K(r)\propto r^{\alpha}$ is empirically
found to hold in the settled halo bodies, with slopes around
$1.25$ (see Navarro et al. 2008). This apparently universal
halo feature allows recasting the pressure $\rho(r)\,
\sigma_r^2(r)\propto K(r)\,\rho^{5/3}(r)$ in terms of the
density, to balance self-gravity for the equilibrium.

As to the latter we have used the isotropic Jeans equation to
derive the `$\alpha$-\emph{profiles}' for DM quantities like
$\rho(r)$ and $\sigma_r^2(r)$, having ascertained that $\alpha$
is to lie within the narrow range $1.25 - 1.3$ from a
state-of-the-art semianalytic study of the cosmogonic halo
development, see Lapi \& Cavaliere (2009, hereafter LC09).

In this \emph{Letter} we first refine these profiles to
conditions of developing outskirts and anisotropic velocity
dispersion; then we test them against recent data that join
observations of strong and weak gravitational lensing (GL).

\begin{figure*}[!t]
\epsscale{0.7}\plotone{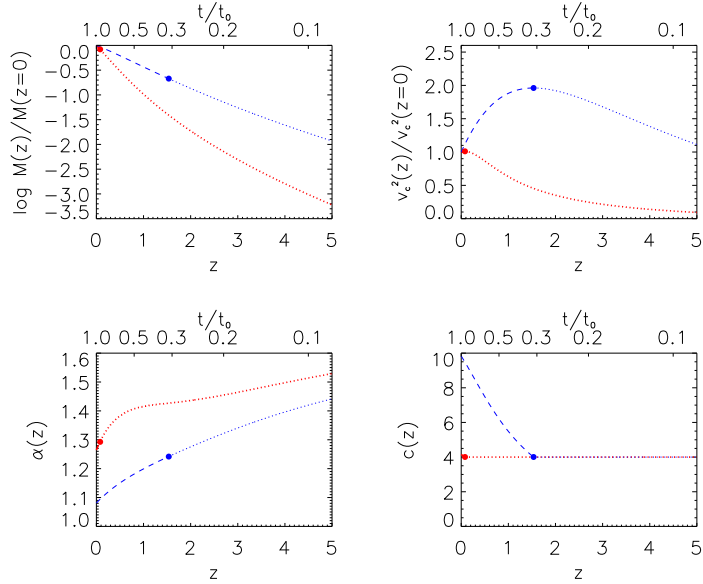}\caption{Evolutionary
track of a halo's mass $M$, circular velocity $v_c^2$, entropy
slope $\alpha$, and concentration $c$ for a current overall
mass of $10^{15}\, M_{\odot}$. In all panels: red lines refer
to the history of an \emph{average} cluster; blue ones refer to
the main progenitor to illustrate the \emph{variance} (see
\S~2.1 for details). Lines are dotted during the fast collapse
of the halo body, and dashed during the slow accretion of the
outskirts; big dots locate the cosmogonic \emph{transition},
when $\alpha=1.29$ and $1.25$ are seen to hold (bottom left
panel) for the average and the main progenitor history,
respectively. During the subsequent slow accretion such values
are retained in the halo body, while they slowly decrease at
the outskirts' boundary following $\alpha(z)$.}
\end{figure*}

\section{Halo development and equilibrium}

Our framework will be provided by the standard
$\Lambda$-cosmology, i.e., a flat Universe with normalized
matter density $\Omega_M = 0.27$, dark energy density
$\Omega_\Lambda = 0.73$, and Hubble constant $H_0 = 72$ km
s$^{-1}$ Mpc$^{-1}$. To bridge the matter- to the dark
energy-dominated era across the cosmological crossover at
$z\approx 0.5$, we shall express the redshift-time relation
$(1+z)\propto t^{-q}$ in terms of the parameter $q$ growing
from $2/3$ to $4/5$.

\subsection{The entropy slope from cosmogonic development}

The evolutions of the current bounding radius $R$, of the
circular velocity $v_c^2$, and of the entropy $K$ are obtained
by LC09 in terms of the halo mass $M$ and its growth rate
$\dot{M}$ from the simple scaling laws $R\propto
M/\dot{M}^{2/3}$, $v_c^2\propto \dot{M}^{2/3}$, and $K\propto
R\, M^{1/3}$. Whence straightforward algebra leads to express
the entropy slope $\alpha\equiv \mathrm{d}\log\,
K/\mathrm{d}\log R$ as
\begin{equation}
\alpha=1+{1\over 1+2\,\epsilon/q}~,
\end{equation}
in terms of the inverse growth rate $\epsilon\equiv -\,
\mathrm{d}\log\, (1+z)/\mathrm{d}\log M= q\, M/\dot{M}\, t$.

With $\epsilon\approx 1$ marking the cosmogonic transition from
fast to slow accretion as gauged on the running Hubble
timescale, it is seen from Eq.~(1) that the range
$\alpha\approx 1.25-1.3$ will apply to average halos that began
their slow accretion in the corresponding interval $z_t\approx
1.5-0.2$. The range of $\alpha$ is narrow as the evolution of
$\epsilon(t)/q(t)$ is \emph{slower} than for $\epsilon(t)$ and
$q(t)$ separately, which explains why closely similar values of
$\alpha$ are found in the halo bodies from different
simulations.

Such a behavior is checked and refined in terms of the detailed
evolution of $\epsilon(t)$; its \emph{average} evolutionary
track is obtained from integrating for $M(t)$ the differential
equation
\begin{equation}
\dot{M}(M,t)=\int_0^{M}{dM'}~(M-M')\,{\mathrm{d}^2\,P_{M'\rightarrow
M}\over \mathrm{d}\,M'\mathrm{d}\,t}~,
\end{equation}
with the state-of-the-art \emph{kernel} detailed in Appendix A
of LC09. We illustrate as red lines in Fig.~1 our outcomes for
an average cluster with current overall mass $M\approx
10^{15}\, M_{\odot}$; we plot the redshift evolution of the
mass $M(z)$, of the circular velocity $v_c(z)$, of the entropy
slope $\alpha(z)$. Note that our approach, which includes
ellipsoidal collapse of the body and outskirts growth
controlled by $\Lambda$, renders the peaked behavior of
$v_c^2(t)$ in remarkable agreement with the detailed
simulations. For $M\approx 10^{15}\, M_{\odot}$ the transition
occurs at $z_t\approx 0.2$, the entropy slope at $z_t$ is
around $\alpha\approx 1.3$, and the outskirts are currently
rudimentary.

But considerable \emph{variance} arises from the stochastic
nature of the individual growth histories. As an example of
variant, early \emph{biased} track we focus on the one
associated with the `main progenitor' that constitutes the main
branch of a merging tree (illustrated, e.g., in Cavaliere \&
Menci 2007, their Fig.~1). Such a history obtains from the same
Eq.~(2), with the lower integration limit replaced by $M/2$;
the results for a \emph{current} mass $M\approx 10^{15}\,
M_{\odot}$ are shown as blue lines in Fig.~1. Relative to the
average, this history features a higher transition redshift
$z_t\approx 1.5$, a less massive body with $M\approx 2\times
10^{14}\, M_{\odot}$, an entropy slope $\alpha\approx 1.25$,
and currently extensive outskirts. We find the occurrence of
such biased halos relative to the average to be $0.125:1$ on
integrating the kernel of Eq.~(2) over the corresponding two
histories.

An imprint of the transition redshift $z_t$ is provided by the
concentration parameter $c(z)$, that in overall terms scales as
$[M(z)/M(z_t)]^{1/3}$; in fact, Zhao et al. (2003) and Wechsler
et al. (2006) describe its increase for $z < z_t$ with the
approximation $c(z)\approx 4\, (1+z_t)/(1+z)$, adopted in the
bottom right panel of our Fig.~1. It is seen that for the
average history the present concentration reads $c\approx 4$,
while for the main progenitor it takes on values $c\approx 10$.

From overall halo development we turn now to profiles for the
equilibrium following the transition time.

\begin{figure*}[!t]
\epsscale{0.7}\plotone{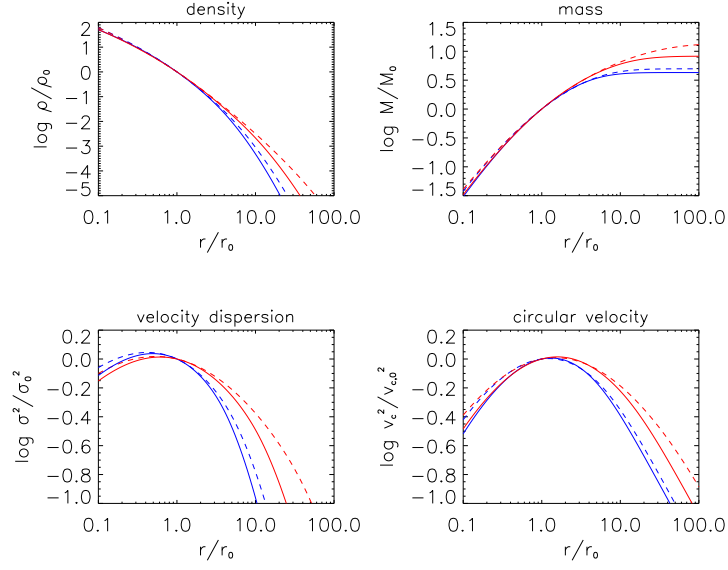}\caption{Radial runs
for the $\alpha$-profiles: density $\rho$, mass $M$, velocity
dispersion $\sigma_r$, and circular velocity $v_c$; the
profiles are normalized to $1$ at the point $r_0$ where
$\gamma=\gamma_0$ holds (see Eq.~4), with the main body
spanning the range $r\la 2\,r_0$. Curves are for $\alpha=1.25$
(blue) and $1.29$ (red), in the isotropic (dashed) and
anisotropic (solid) equilibria; we adopt the anisotropy given
in Eq.~(5) with $\beta(0)=-0.1$ and $\beta'=0.2$, the latter
parameter being actually irrelevant (see \S~3.1).}
\end{figure*}

\subsection{$\alpha$-profiles from Jeans equilibrium}

Physical profiles are derived from the above values of $\alpha$
inserted into the radial Jeans equation, with pressure
$\rho\sigma^2_r \propto r^{\alpha}\, \rho^{5/3}$ and anisotropy
described in terms of the standard Binney (1978) parameter
$\beta\equiv 1-\sigma_\theta^2/\sigma_r^2$. In terms of the
density slope $\gamma\equiv -\mathrm{d}\log\rho/\mathrm{d}\log
r$ Jeans may be recast to read
\begin{equation}
\gamma = {3\over 5}\,(\alpha+{v_c^2\over \sigma_r^2}) + {6\over
5}\,\beta~.
\end{equation}
When supplemented with the mass definition $M(<r)\equiv
4\pi\int_0^r{\mathrm{d}r'}~r'^2\,\rho(r')$ entering $v_c^2$,
this constitutes an integro-differential equation for
$\rho(r)$, that by double differentiation reduces to a handy
$2^{\rm nd}$ order differential equation for $\gamma$ (Austin
et al. 2005, Dehnen \& McLaughlin 2005).

With $\alpha =$ const and $\beta = 0$ (meaning isotropy), LC09
found that physical solutions, that we named
'$\alpha$-profiles', exist for $\alpha\leq 35/27 =
1.\overline{296}$; the corresponding density runs steepen
monotonically outwards and satisfy physical central and outer
boundary conditions, respectively: a round minimum of the
potential along with a round maximum of the pressure; a finite
(hence definite) overall mass. In Fig.~2 we report as dashed
lines the $\alpha$-\emph{profiles} for various quantities:
density $\rho(r)$, mass $M(<r)$, circular velocity $v_c^2(r)$,
and velocity dispersion $\sigma_r^2(r)$.

The behavior of the density run (top left panel of Fig.~2) is
highlighted by the analytic expressions of the slopes
\begin{equation}
\gamma_a \equiv {3\over 5}\,\alpha~,~~~~ \gamma_0\equiv
6-3\alpha~,~~~~ \gamma_b\equiv {3\over2}\,(1+\alpha)~.
\end{equation}
These start with the central ($r\rightarrow 0$) value
$\gamma_a\approx 0.75-078$, progressively steepen to
$\gamma_0\approx 2.25-2.1$ at the point $r_0$ that marks the
halo main body, and steepen further into the outskirts to the
value $\gamma_b\approx 3.38-3.44$.

Monotonic behavior and physical boundary conditions for the
$\alpha$-profiles are seen (cf. LC09) to imply a maximal value
$\kappa_{\mathrm{crit}}(\alpha) = v^2_c/\sigma^2\approx
2.6-2.5$ in the body, at the point $r_p\ga r_0$ where
$v_c^2(r)$ peaks (see Fig.~2, bottom right panel); there the
slope reads $\gamma_p = 3\,(\alpha+k_{\rm crit})/5\approx 2.32
- 2.28$ after Eq.~(3).

Thus the inner slopes are considerably \emph{flatter} as to
yield a smooth central pressure, while the outer one is
\emph{steeper} as to yield a definite overall mass, compared to
the empirical NFW formula (Navarro et al. 1997). The latter, in
fact, implies infinite mass, and angled central pressure and
potential.

The concentration parameter for the $\alpha$-profiles is given
in detail by $c\equiv R/r_{-2}$ in terms of the radius $r_{-2}$
where $\gamma=2$. This may be viewed as a measure of central
condensation (small $r_{-2}$) and/or outskirts' extension
(large $R$).

\begin{figure*}[!t]
\nonumber\epsscale{0.7}\plotone{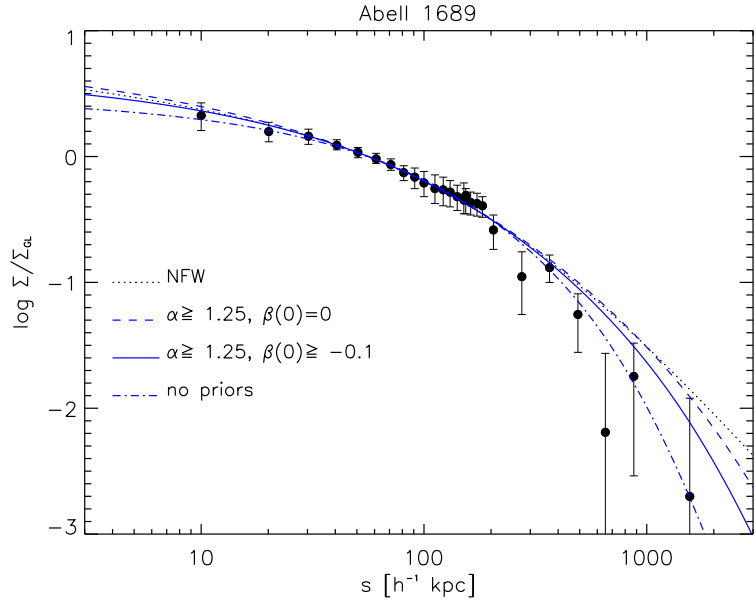}\caption{Surface
density runs for the cluster A1689. Filled symbols represent
the data by Lemze et al. (2008; see also Umetsu \& Broadhurst
2008) from joint strong and weak GL observations. Blue lines
illustrate fits to the data from our $\alpha$-profiles with no
prior (dot-dashed), or with priors: $\alpha\geq 1.25$ and
$\beta(0) = 0$ (dashed); $\alpha\geq 1.25$ and $\beta(0)\geq
-0.1$ (solid). For comparison, the black dotted line shows the
NFW fit. We report in Table 1 the values of the fitting
parameters $\alpha$, $\beta(0)$, and $c$ and of the
corresponding minimum $\chi^2/dof$.}
\end{figure*}

\section{Cluster profiles tested with GL observations}

A significant observational test requires to go \emph{beyond}
the limitations to $\alpha =$ const and isotropy; we tackle
these issues in turn.

\subsection{Refined $\alpha$-profiles}

As to the latter, here we include on the r.h.s. of Eq.~(3) the
anisotropy term. This clearly will \emph{steepen} the density
run for positive $\beta$ meaning radial dominance, as expected
in the outskirts from infalling cold matter. Tangential
components must develop toward the center, as expected from
increasing importance of angular momentum effects (see LC09),
and as supported by numerical simulations (Austin et al. 2005;
Hansen \& Moore 2005; Dehnen \& McLaughlin 2005). In detail,
the latter suggest the effective linear approximation
\begin{equation}
\beta(r)\approx \beta(0)+\beta'\,[\gamma(r)-\gamma_a]
\end{equation}
with $\beta(0)\ga -0.1$ and $\beta'\approx 0.2$, limited to
$\beta(r)\la 0.5$. We find (Fig.~2) the corresponding $\rho(r)$
to be slightly \emph{flattened} at the center by a weakly
negative $\beta(0)$, and considerably \emph{steepened} into the
outskirts where $\beta(r)$ grows substantially positive.
Specifically, the following simple rules apply: the slope
$\beta'$ in Eq.~(5) is found to drop out from the derivatives
of the Jeans equation (Dehnen \& McLaughlin 2005); the upper
bound to $\alpha$ now reads
$\tilde{\alpha}=35/27-4\beta(0)/27$; $\gamma_a$ is modified
into $\gamma_a = 3\alpha/5+6\beta(0)/5$ while $\gamma_0$ and
$\gamma_b$ retain their form.

As to the former and minor issue, we include the slowly
decreasing run $\alpha(r)$ (cf. Fig.~1) enforced within
outskirts developing by slow mass accretion, as the outer scale
is stretched out while the body stays put. Clearly, this
affects little the inner profile of $\rho(r)$ as Jeans itself
(with its inner boundary conditions) works from the inside out,
but it tilts \emph{down} $\rho(r)$ appreciably into the
outskirts.

In Fig.~2 we represent as solid lines the $\alpha$-profiles
\emph{refined} as to include both the declining $\alpha(r)$ and
the anisotropies discussed above.

\subsection{Testing the $\alpha$-profiles and their development}

Now we turn to testing these refined profiles against the
recent, extensive GL observations of the cluster A1689 that
join strong and weak lensing to cover scales from $10^{-2}$ to
$2.1$ Mpc, the latter being the virial radius $R$, see
Broadhurst et al. (2005), Halkola et al. (2006), Limousin et
al. (2007), Umetsu \& Broadhurst (2008); we focus on the
dataset presented by the latter authors and refined in Lemze et
al. (2008).

These observations concern the surface density, for which our
benchmark is constituted by the refined $\alpha$-profiles
integrated over the l.o.s. at a projected distance $s$ from the
center
\begin{equation}
\Sigma(s) = 2\, \int_s^{R}{\mathrm{d}r}\, {r\, \rho(r)\over
\sqrt{r^2-s^2}}~.
\end{equation}
In Fig.~3 we illustrate as blue lines the outcomes of fits to
the data in terms of our refined $\alpha$-profiles; for the
anisotropy we have adopted Eq.~(5), with $\beta(0)$ as the only
effective parameter, see \S~3.1.

The dashed line represents the best fit for isotropic profiles
($\beta = 0$) with the prior $\alpha\geq 1.25$ from the
two-stage development (see \S~2.1); the solid line refers to
the best fit for anisotropic profiles subject to the priors
$\alpha\geq 1.25$ and $\beta(0)\geq -0.1$ from simulations (see
\S~3.1); the dot-dashed line refers to the best fit for
anisotropic profiles with no prior. For comparison, the black
dotted line illustrates the fit with the NFW formula.

In Table 1 we report the corresponding values of the fitting
parameters $\alpha$, $\beta(0)$, and $c$ (with their $68\%$
uncertainty), and of the corresponding minimum $\chi^2/dof$.
The physical $\alpha$-profiles generally provide fits of
comparable or better quality than the empirical NFW formula;
note that an optimal fit obtains with no priors on $\alpha$ and
$\beta$, but at the cost of ignoring the information on the
former as provided by the two-stage development, and on the
latter as provided by the numerical simulations.

\begin{deluxetable}{lcccc}
\tabletypesize{} \tablecaption{Results of the fits to the
surface density of A1689} \tablewidth{0pt}
\tablehead{\colhead{Reference in Fig.~3} & \colhead{$\alpha$} &
\colhead{$\beta(0)$} & \colhead{$c$} & \colhead{$\chi^2/dof$}}
\startdata
\\
NFW & $---$ & $---$ & $12.2^{+3.2}_{-2.7}$ & $7.94/(26-2)$\\
\\
$\alpha\geq 1.25$, $\beta(0)=0$ & $1.25^{+0.03}$ & $0$ & $10.7^{+3.2}_{-2.5}$ & $8.78/(26-3)$\\
\\
$\alpha\geq 1.25$, $\beta(0)\geq -0.1$ & $1.25^{+0.04}$ & $-0.1^{+0.09}$ & $11.6^{+2.8}_{-2.4}$ & $6.55/(26-4)$\\
\\
no priors & $1.18^{+0.09}_{-0.13}$ & $-0.15^{+0.23}_{-0.21}$ & $12.4^{+2.2}_{-2.1}$ & $4.13/(26-4)$\\
\enddata
\tablecomments{For the anisotropy we have adopted Eq.~(5),
where $\beta(0)$ is the only relevant free parameter, see
\S~3.1.}
\end{deluxetable}

We find that \emph{balanced} fits to the surface density in
A1689 require $c\approx 10$; lower values would cause wide
overshooting of the outer points, while higher ones would allow
approaching these (with little impact on $\chi^2$ owing to the
large uncertainties) at the cost of overshooting the precise
inner points.

A similar balancing has been found by Broadhurst et al. (2008)
from fits with the empirical NFW formula, yielding generally
high concentrations. These authors suggest that large values of
$c$ may be understood in terms of formation redshifts earlier
than expected from the standard $\Lambda$CDM cosmogony, as
quantitatively focused by Sadeh \& Rephaeli (2008). From the
perspective of our $\alpha$-profiles, concentrations $c\approx
10$ are strictly related to transition redshifts $z_t\approx
1.5$ based on the state-of-the-art evolution of biased
$\Lambda$CDM halos, see Eq.~(2) and Fig.~1; on the same ground,
we compute their occurrence to be bounded by about $13\%$ in
blind sampling.

On the other hand, such early biased halos tend to be favored
with GL data. This is because strong GL observations are
favored by centrally flat profiles and steep outskirts
producing conspicuously large Einstein rings (Broadhurst \&
Barkana 2008). Meanwhile, weak GL observations require
extensive outskirts to affect numerous background galaxies.

For A1689 the latter data with their uncertainties are not yet
sharply constraining the fit, but we expect convergence toward
the physical profiles as the uncertainties are narrowed down by
improved control over the redshift distribution of background
galaxies (see Medezinski et al. 2007, Limousin et al. 2007,
Umetsu \& Broadhurst 2008). On the other hand, more and more
clusters are being covered by GL observations, which often find
evidence of centrally flat density runs (Brada\v{c} et al.
2008, Sand et al. 2008, Richard et al. 2009, Oguri et al.
2009), consistently with our physical $\alpha$-profiles.

\section{Discussion and conclusions}

To sum up, the DM halo benchmark we test with GL observations
is comprised of a time and a space behavior, that we strictly
link in the framework of $\Lambda$CDM cosmogony. As to time, we
find a \emph{narrow} range of the DM entropy slope $\alpha$ as
the outcome of a \emph{two-stage} development comprising an
early fast collapse of the body followed by a slow, inside-out
growth of the outskirts.

As to space, the physical $\alpha$-profiles we derive feature
density runs $\rho(r)$ intrinsically \emph{flatter} toward the
center, and intrinsically \emph{steeper} toward the outskirts
as to yield a definite mass, relative to the singular NFW
rendition of early $N$-body data. We find these runs to be
stable with, or even \emph{sharpened} by anisotropy.

These physical $\alpha$-profiles improve at both small and
large scales the fits to the GL data, including the extensively
probed case of A1689. Here the present analysis requires a halo
\emph{biased} toward a main progenitor lineage, with
non-standard concentration $c\approx 10$ marking a body
collapsed early at $z_t\approx 1.5$ and late extensive
outskirts; we find such halos to comprise some $10\%$ of the
clusters.

Alternative proposals to relieve the tension of the GL
observations with the profiles expected for average clusters
include density cores flattened by degenerate pressure
(Nakajima \& Morikawa 2007), or concentrations enhanced by some
$30\%$ owing to sharply prolate triaxialities (see Hennawi et
al. 2007; Oguri \& Blandford 2009; Corless et al. 2009) which
on the other hand would produce steeper central slopes. The
$\alpha$-profiles and the two-stage development dispense with
such contrasting interpretations, providing sharp and
consistent shapes linked with early-biased histories. This view
clearly invites blind sampling of more clusters in GL.

On the other hand, as we discuss below, X-ray data will provide
an \emph{independent} line of evidence concerning profiles and
concentrations. This is based on the other major component of
clusters, i.e., the intracluster plasma (ICP) which settles to
its own equilibrium within the DM potential well, and emits
strong X rays by thermal bremsstrahlung (see Sarazin 1988).

ICP information from spectroscopy (yielding the temperature
$T$) and X-ray brightness (yielding the squared number density
$n^2$) are best combined in the form of the thermodynamic ICP
entropy (adiabat) $k(r)\equiv k_B T/n^{2/3}$; this modulates
the ICP equilibrium within the DM potential well, and
throughout the cluster body follows a powerlaw run $k(r)\propto
r^a$ with $a\la 1.1$.

In fact, Lapi et al. (2005) and Cavaliere et al. (2009) compute
the slope $a$ to be expected in the cluster outskirts from
accretion of external gas shocked at about the virial $R$ (see
also Tozzi \& Norman 2001). They find $a\approx
2.37-0.71/\Delta\phi(c)$ in terms of the potential drop
$v_c^2(R)\,\Delta\phi(c)$ from the turnaround to $R$;
meanwhile, the ICP density follows $n(r)\propto r^{-g}$ with
$g\approx 1.42+0.47/\Delta\phi(c)$. Values $\Delta\phi\approx
0.56$ are seen to apply for $\alpha$-profiles with
concentration $c\approx 5$, to yield $a\approx 1.1$ and
$g\approx 2.2$ as measured in many clusters. But in clusters
with extensive outskirts and higher concentrations the outer
potential is shallower and $\Delta\phi(c)$ smaller; when
$c\approx 10$ one finds $\Delta\phi\approx 0.47$, yielding an
intrinsically \emph{flatter} $a\approx 0.85$ (and a steeper
$g\approx 2.4$ in the absence of large central energy
discharges). This theoretical expectation finds gratifying
support in the flat $a\approx 0.8$ observed in A1689 by Lemze
et al. (2008).

We note that the $\alpha$-profiles provide a physical
\emph{benchmark} also useful in the context of probing cold DM
annihilations [$\propto \rho^2(r)$] or decays [$\propto
\rho(r)$] through diffuse $\gamma$-ray emissions expected from
the Galaxy center (e.g., Bertone et al. 2008), and the positron
excess recently detected by \textsl{PAMELA} (e.g., Adriani et
al. 2008). We will investigate the issue elsewhere.

Finally, since a two-stage development also applies to hot DM
cosmogonies (though with $c$ decreasing in cosmic time, see
Wang \& White 2008), we comment upon the case for cold DM on
the basis of the radial run $\sigma^2_r(r) \propto
\rho^{2/3}(r)\, K(r)\propto r^{\alpha-2\gamma(r)/3}$. We expect
that \emph{cold} DM halos will be marked by $\sigma_r^2(r)$
falling down to a few hundreds km s$^{-1}$ into the outskirts,
cf. Fig.~2 with the dynamical observations by Lemze et al.
(2009, their Fig.~7). Such a behavior will provide evidence for
a truly cold nature of the DM.

\begin{acknowledgements}
Work partially supported by Agenzia Spaziale Italiana (ASI). We
thank T. Broadhurst, R. Fusco-Femiano, E. Medezinski, P.
Natoli, and K. Umetsu for informative and useful discussions.
We acknowledge an anonymous referee for constructive and
helpful comments.
\end{acknowledgements}

\end{document}